\documentclass[a4paper,fleqn,usenatbib]{mnras}

\usepackage{newtxtext,newtxmath}

\usepackage[T1]{fontenc}
\usepackage{ae,aecompl}
\usepackage[normalem]{ulem}

\usepackage{graphicx}	
\usepackage{amsmath}	
\usepackage{amssymb}	

\def\mnras{MNRAS}
\def\prd{Phys Rev D}
\def\physrep{Physics Reports}
\def\nat{Nature}
\def\nar{New Astronomy Reviews}
\def\aap{A\&A}
\def\apj{ApJ}
\def\apjs{ApJS}
\def\apjl{ApJL}
\def\aj{AJ}

\def\lsim{\mathrel{\lower0.6ex\hbox{$\buildrel {\textstyle <}
 \over {\scriptstyle \sim}$}}}
\def\gsim{\mathrel{\lower0.6ex\hbox{$\buildrel {\textstyle >}
 \over {\scriptstyle \sim}$}}}

\def\racc{$\bf r_{\rm acc}$}
\def\eone{${\bf e}_{1}$}
\def\etwo{${\bf e}_{2}$}
\def\ethree{${\bf e}_{3}$}

\title[Universal subhalo accretion in $\Lambda$CDM and $\Lambda$WDM]{Universal subhalo accretion in cold and warm dark matter cosmologies}

\author[Kubik et al]
{Bogna Kubik$^{1}$, Noam I. Libeskind$^{2}$, Alexander Knebe$^{3,4}$, H\'el\`ene Courtois$^{1}$,
\newauthor Gustavo Yepes$^{3,4}$, Stefan Gottl\"ober$^{2}$, Yehuda Hoffman$^{5}$\\
$^1$University of Lyon, UCB Lyon 1/CNRS/IN2P3; IPN Lyon, France\\
$^2$Leibniz-Institut f\"ur Astrophysik Potsdam (AIP), An der Sternwarte 16, D-14482 Potsdam, Germany \\
$^3$Departamento de F\'{\i}sica Te\'orica M8, Universidad Aut\'onoma de Madrid (UAM), Madrid, Spain\\
$^4$Astro UAM, UAM, Unidad Asociada CSIC.\\
$^5$Racah Institute of Physics, Hebrew University, Jerusalem 91904, Israel\\ 
}

\date{Accepted XXX. Received YYY; in original form ZZZ}

\pubyear{2016}

\begin{document}
\label{firstpage}
\pagerange{\pageref{firstpage}--\pageref{lastpage}}
\maketitle

 \begin{abstract}
The influence of the large scale structure on host halos may be studied by examining the angular infall pattern of subhalos. In particular, since warm and cold dark matter cosmologies predict different abundances and internal properties for halos at the low mass end of the mass function, it is interesting to examine if there are differences in how these low mass halos are accreted. The accretion events are defined as the moment a halo becomes a substructure, namely when it crosses its host's virial radius. We quantify the cosmic web at each point by the shear tensor and examine where, with respect to its eigenvectors, such accretion events occur in cold ($\Lambda$CDM) and warm (1keV sterile neutrino WDM) dark matter cosmological models. We find that the CDM and WDM subhalos are preferentially accreted along the principal axis of the shear tensor corresponding to the direction of weakest collapse. The beaming strength is modulated by the host and subhalo masses and by the redshift at which the accretion event 
occurs. 
Although strongest for the most massive hosts and subhalos at high redshift, the preferential infall is found to be always aligned with the axis of weakest collapse, thus we say that it has universal nature. We compare the strength of beaming in the WDM cosmology with the one found in the $\Lambda$CDM scenario. While the main findings remain the same, the accretion in the WDM model for the most massive host halos appears more beamed than in $\Lambda$CDM cosmology across all the redshifts.
\end{abstract}




\section{Introduction}
\label{section:intro} 
The nature of dark matter remains one of the central unsolved problems in modern cosmology and particle physics.
Data from the cosmic microwave background indicate that the dark matter is made up of nonbaryonic elementary particles \citep[e.g.][]{2011ApJS..192...16L}, but exactly which kind \citep[or kinds,][]{2009PhRvD..79k5002Z, 2012PhRvD..85h3523D, 2012PhRvD..86g6015A, 2014PhRvL.113g1303M, 2015PhRvD..91i5006G} of particles are involved is not yet known.

For the past 30 years the attention has focused on the cold dark matter (CDM) model \citep{1984Natur.311..517B,1985ApJ...292..371D,1985Natur.317..595F}, for which there are well motivated particle candidates, for example the lightest supersymmetric particle, neutralino \citep{1984NuPhB.238..453E}  or the axion \citep{1983PhLB..120..127P}.

In the CDM paradigm, dark matter particles possess a negligible initial velocity dispersion and structure formation proceeds in a strictly hierarchical bottom-up manner \citep[e.g.][]{1978MNRAS.183..341W}. Small perturbations imprinted on the primordial density field grow via gravitational instabilities and merge with each other to create larger halos as well as the complex structures we observe today.

On large scales, the $\Lambda$CDM model is consistent with the cosmic microwave background (CMB) anisotropy spectrum measured by the Wilkinson Microwave Anisotropy Probe \cite[WMAP,][]{2013ApJS..208...19H} and PLANCK \citep{2016A&A...594A..13P} as well as with observations of the large-scale $(k \leq 0.1 h$Mpc${}^{-1}$) galaxy clustering spectrum measured by the Sloan Digital Sky Survey \citep[SDSS,][]{2006PhRvD..74l3507T}. Moreover, $\Lambda$CDM has significant predictive power because the only parameter describing the dark matter in this theory, $\Omega_m$, is now very well constrained by observational data \citep[e.g.][]{2016A&A...594A..13P}. 

However, there exist a number of observations that may indicate a lack of small scale perturbations in the initial density field. For example, it has long been known that $\Lambda$CDM theory predicts many more small mass halos than the number of dwarf galaxies that we see around the Milky Way \citep{1993MNRAS.264..201K, 1999ApJ...522...82K, 1999ApJ...524L..19M, 2010arXiv1009.4505B}.

A related issue is the so-called "too big to fail" (TBTF) problem which states that many of the satellites predicted by $\Lambda$CDM cosmology are simply too massive to have star formation suppressed by reionization processes \cite{2011MNRAS.415L..40B, 2012MNRAS.422.1203B} while no galaxies that could match the kinematics of these massive haloes were observed. A lack of visible satellites was reported at the scale of the Milky Way and Andromeda \citep{2011MNRAS.415L..40B, 2012MNRAS.422.1203B, 2014MNRAS.440.3511T, 2015ApJ...799L..13C} or at the scale of Local Group for the field galaxies \citep{2012MNRAS.425.2817F, 2014MNRAS.444..222G, 2015A&A...574A.113P, 2015MNRAS.454.1798K}. Based on statistical arguments \citet{2013ApJ...767...92R} further suggested that the TBTF problem should be generically present for MW-sized galaxies. 

Finally, the presence of a cusp in the centers of $\Lambda$CDM halos, one of the earliest and strongest results derived from cosmological N-body simulations \citep{1991ApJ...378..496D}, for which the logarithmic slope $\alpha = d \ln \rho/d\ln r $ $\sim-1$ at small radii \citep{1996ApJ...462..563N, 1997ApJ...490..493N} seems to be inconsistent with predictions of the density profiles in the $\Lambda$CDM \citep{1994Natur.370..629M, 2001ApJ...552L..23D, 2005ApJ...621..757S, 2010AdAst2010E...5D}.

The discussions are still ongoing though. The "missing satellite problem" is partially alleviated by many faint MW satellites that have been discovered in the SDSS \citep{2005AJ....129.2692W, 2007ApJ...654..897B} and in DES \citep{2015ApJ...807...50B, 2015ApJ...813..109D}. It also has become clear that up to a factor of $\sim$5-20 times as many faint galaxies could remain undetected at present owing to incomplete sky coverage \citep{2008ApJ...688..277T}, luminosity bias \citep{2009AJ....137..450W} or surface brightness limits \citep{2010ApJ...717.1043B}. Another possible solution arises if the lowest mass dark matter haloes are inefficient at forming stars due to the early reionization of the intergalactic medium \citep{2000ApJ...539..517B, 2002ApJ...572L..23S, 2002MNRAS.333..177B}. 

The attempts of facing the "too big to fail" problem  include suggestions to change the nature of the dark matter from 'cold' to 'self-interacting' \citep[e.g.][]{2000PhRvL..84.3760S, 2012MNRAS.423.3740V, 2017MNRAS.469.2845D,2017arXiv170502358T} which forms more cored and less dense halos, a possible scatter from the stochastic nature of structure formation so that a fraction of the more massive subhaloes remain dark \citep[e.g.][]{2012JCAP...12..007P, 2013ApJ...767...92R, 2013ApJ...773..172R}, lowering the mass of the MW host halo \citep[e.g.][]{2012MNRAS.424.2715W,2013MNRAS.428.1696V} to better fit the observed MW's velocity function or tidal interactions \citep{2017MNRAS.468.4887W, 2016MNRAS.457.1931S}  that reduce the dark matter content of several Milky Way dwarf satellites. Another possible common solution to both the "cusp-core" and the "too big to fail" problem is the formation of cores through the effects of baryonic feedbacks driven by the supernovae gas outflows  \citep[e.g.][]{2012MNRAS.422.1231G,
2013MNRAS.433.3539G,2014MNRAS.437..415D,2014ApJ...789L..17M,2014ApJ...784L..14B,2015MNRAS.450.3920B}.

Warm dark matter (WDM) has been suggested as another alternative in dealing with many of these issues \citep{2010MNRAS.404L..16M, 2012MNRAS.420.2318L, 2014MNRAS.437.2922P}. One of the most promising WDM candidates is a sterile (right-handed) neutrino which solves, besides problems related to structure formation, a number of other open questions such as the observed neutrino oscillation rates and baryogenesis (see \citet{2005PhLB..620...17A} and \citet{2009PhR...481....1K} for alternatives to the neutrino minimal standard model described in \citet{2009ARNPS..59..191B} ). 

WDM particles are relativistic when they decouple from the primordial plasma and become non-relativistic during the radiation-dominated era. This causes the particles to free stream out of small perturbations, giving rise to a cutoff in the linear matter power spectrum and an associated suppression of structure formation on small scales \citep{2000ApJ...542..622C,2001ApJ...559..516A,2001ApJ...556...93B,2002MNRAS.329..813K}.

It has been suggested that a free-streaming length corresponding to that of a thermal relic WDM particles such as sterile neutrinos or gravitinos, with a mass of 1-2 keV provides better agreement between the most recent data and numerical simulations \citep{2013MNRAS.432.3218D}. The present tightest limits on the dark matter coldness are placed by the Intergalactic Medium (IGM), namely the Lyman-$\alpha$ forest flux power spectrum as seen in absorption in quasar spectra, and give a lower limit of WDM mass of 3 keV \citep{2005PhRvD..71f3534V, 2006PhRvD..73f3513A, 2009JCAP...05..012B, 2013PhRvD..88d3502V}. These values are larger than those typically chosen to reconcile local properties of the satellite populations of the Milky Way which are in the range 1.5 - 2 keV \citep{2012MNRAS.420.2318L, 2014PhRvL.112p1303A}.

Since WDM and CDM have different low mass halo populations, but (by design) more or less identical large scale properties, we are interested in the dark matter (DM) behavior on the quasi-linear scales along which subhalos are accreted. Specifically, we wish to understand the anisotropic infall pattern of DM halos with respect to the large scale structure in these two cosmological models. Isotropy on large scales is one of the fundamental assumptions in modern cosmology and is widely verified in large galaxy surveys. However on scales of a few Mpc, the matter distribution is structured in clusters and filaments.

The notion of anisotropy in the context of structure formation was first introduced by Zel'dovich (\citealt{1982Natur.300..407Z}). Since that time the issue of anisotropy down to galactic and cluster scales has been studied. Now observational studies \citep[e.g.][]{2002ASPC..268...81P, 2003ApJ...590L...9K, 2009RAA.....9...41F} and numerical investigations \citep[e.g.][]{2002A&A...395....1F, 2004MNRAS.352..376A,2013MNRAS.428.2489L,2015MNRAS.454.2736C} confirm that galaxies tend to be aligned with their neighbors supporting the vision of the anisotropic infall pattern of the accreting material along filamentary structures \citep[e.g.][]{1997MNRAS.290..411T,2004ApJ...603....7K}. Indeed, even on the smallest scales there has been evidence that satellites (or planes of satellites) within larger halos are influenced by the filamentary network of the cosmic web \citep{2005MNRAS.363..146L,2015MNRAS.452.1052L,2016ApJ...829...58G}.

The attempt to understand and relate the anisotropic nature of the mass assembly of halos and galaxies to the inherent anisotropic nature of the cosmic web calls for a classification of the LSS that can be easily defined in a scale free way across different mass scales and different redshifts. 

A dynamic classification of the cosmic web, based on counting the number of positive eigenvalues of the tidal tensor, i.e. the Hessian of the gravitational potential, was proposed by \citet{2007MNRAS.375..489H} and extended by \cite{2009MNRAS.396.1815F}. This approach is driven by the Zeldovich approximation and its web classification agrees remarkably well with the visual impression one has in viewing the large scale structure on linear and quasi-linear scales, i.e. scales larger than a few Mpcs. However, in going down to smaller scales this approach fails to recover the fine web exhibited by high resolution numerical simulations. 

Guided by the fact that in the linear regime the velocity and density fields are essentially proportional to each other \citep{2012MNRAS.425.2049H} reformulated the classification of \citet{2007MNRAS.375..489H} in terms of the velocity shear tensor. They showed that the web defined by the velocity shear resolves much smaller structures than the web defined by the tidal field. The two possible definitions of the web coincide on large scales, where the linear regime prevails, but they depart with going to small scales where the non-linear dynamics manifests itself.

\citet{2014MNRAS.441.1974L} have recently shown that the principal directions of the shear tensor remain coherent over a wide range of redshifts and spatial scales in $\Lambda$CDM cosmology. Furthermore \citet{2014MNRAS.443.1274L} examined the beaming of subhalos, with respect to shear field, in a $\Lambda$CDM simulation and found a convincing alignment signal between accretion and the axis of slowest material compression on the quasi-linear scales surrounding halos. The beaming effect was termed ``universal'' as it occurs -- with varying degrees of strength -- for all hosts, all accretion events, at all times, in all environments and along a multitude of scales. Such a beaming thus appears to be a generic feature of a $\Lambda$CDM universe and is likely related to anisotropies such as planes of satellites \citep{2013MNRAS.435.2116P,2015ApJ...802L..25T}.

In this work we extend the study of \citet{2014MNRAS.443.1274L} to the $\Lambda$WDM cosmology and examine the preferential infall pattern of WDM halos with respect to the large scale structure down to scales of few hundreds of kpc. Our aim is to quantify and, as far as possible, to interpret, the differences that emerge between the two cosmologies.

\section{Method}
\label{section:method}

\subsection{Simulations}
\label{subsection:simulations}
We analyze two DM-only N-body simulations: i) cold dark matter; ii) warm dark matter with a particle mass of 1 keV corresponding to a sterile neutrino. These simulations have been previously used by: \cite{2009ApJ...700.1779Z,2009MNRAS.399.1611T,2013PASA...30...39L,2014NewAR..58....1Y}.

The simulations start at redshift $z = 100$ in a $L = 64 h^{-1}$ Mpc side box sampled with $1024^3$ particles. The construction of initial conditions for the constrained simulations used in this paper has been described in detail in \cite{2009ApJ...700.1779Z,2010MNRAS.402.1899K,2010MNRAS.401.1889L,2014NewAR..58....1Y} and references therein. In both scenarios the matter and $\Lambda$ density parameters at the present time are chosen accordingly to the cosmic microwave background data WMAP3 (\citealt{2007ApJS..170..377S}):  $\Omega_{\Lambda}$ = 0.76, $\Omega_m$ = 0.24, the expansion rate is $H_0$ = 73km/s/Mpc. The rms mass fluctuation in spheres of 8 Mpc is $\sigma_8$ = 0.75 and $n = 0.95$ is the slope of the power spectrum. Such a simulation gives a mass resolution of $1.89\times 10^7h^{-1}\textrm{M}_\odot$ and we used a spatial softening length of  $1 h^{-1}$ kpc.  135 snapshots, equally spaced in expansion factor, are stored in the $\Lambda$CDM simulation ($z$ = 0 to 11.3) and 190 snapshots 
are generated for $\
Lambda$WDM ($z$ = 0 to 33).

Halos and sub-halos are identified by means of the publicly available halo finder AHF\footnote{Code is publicly avaiable at: \url{http://popia.ft.uam.es/AHF}} (\citealt{2004MNRAS.351..399G, 2009ApJS..182..608K}). AHF locates local over-densities in an adaptatively smoothed density field as prospectives for halo centers. The potential minimum of each density peak is then calculated. Bound particles are then associated to dark matter halos.

In the $\Lambda$WDM simulation numerical fragmentation of halos along filaments is an issue (\citealt{2003MNRAS.345.1285K,2007MNRAS.380...93W,2013MNRAS.434.3337A,2016MNRAS.455.1115H}). In order to protect our analysis against these artificially formed halos we only consider halos larger than a limiting mass $M_{\textrm{lim}} = 10.1 \bar{\rho}dk^{-2}_{\textrm{peak}}$, where $d$ is the mean inter-particle separation, $k_{\textrm{peak}}$ is the wavenumber for which $\Delta^2(k) = k^3P(k)$ reaches its maximum, and $\bar{\rho}=\Omega_{\textrm{m}}\rho_{\textrm{crit}}$ is the mean density, as suggested by \cite{2007MNRAS.380...93W}. In our case $M_{\textrm{lim}} = 3\times10^9h^{-1}\textrm{M}_\odot$ and marks the mass limit below which there is an artificial rise in the $\Lambda$WDM mass function. In order to compare like with like we apply this mass limit to both $\Lambda$CDM and $\Lambda$WDM simulations, ignoring smaller halos.

Accretion events are found by identifying which sub-halos at a given snapshot $z_1$ are identified as "field" halos at the previous snapshot $z_2$ (where $z_1 < z_2$), by building a merger-tree. The accreted position is assumed to be the midpoint, namely a linear interpolation, between the sub-halos positions with respect to the host at $z_1$ and $z_2$.

A small fraction of the accreted sub-halos enter, exit and then re-enter the halo more than once. This multiple-entry needs to be properly subtracted. This is done by tracking all the accreted subhalos back in time through their merger-tree and checking if they were ever identified as subhalos at any previous redshift. For the small fraction of subhalos where this phenomena occurs, only the first entry, i.e. at highest redshift, is considered for the statistical analysis presented here.

\subsection{Velocity shear tensor as a scale independent eigenframe}
\label{subsection:velocity_shear_tensor}

In order to compute the shear tensor at each point in the simulation, the velocity field is gridded according to a Clouds-In-Cell (CIC) scheme. The size of the CIC used here is $256^3$ which for a 64$h^{-1}$Mpc box results in grid cells 0.25$h^{-1}$Mpc per side. These dimensions are chosen such that every mesh cell contains at least one particle at $z = 0$. The motivation behind it is that a mesh which meets this requirement at $z = 0$ will meet it at all redshifts.

The resulting velocity field is then Fourier transformed into $k$-space and smoothed with a Gaussian kernel in order to wash out artificial effects introduced by the preferential axes of the CIC's Cartesian grid. The Gaussian smoothing also sets the scale on which the shear is computed. When a small smoothing kernel is used, non linear features of the velocity field might become apparent. When a large smoothing is used, the field is washed into the linear regime. The size of the CIC cells sets the smallest scale we may probe (i.e 0.25$h^{-1}$ Mpc),  while in principle the largest scale is set simply by the simulation's box size.

The width of the gaussian smoothing is adaptive and depends on the mass of the halo we wish to examine. At each redshift the host halos are divided into five mass bins from $10^9$ to $10^{14}h^{-1}M_{\odot}$, each a decade wide. The median virial radius $R_{\textrm{virmed}}$ for each mass bin is then computed. For each halo we choose to employ the shear field smoothed on 4, 8, and 16 times the median virial radius of the mass bin the halo is in.

Using the Fourier Transform of the velocity field the normalized shear tensor is calculated as:
\begin{equation}
    \Sigma_{ij} = -\frac{1} {2H(z)} \left( \frac {\partial v_i}{\partial r_j} + \frac {\partial v_j}{\partial r_i} \right),
\end{equation}
where $i,j = x, y, z$. The shear tensor is scaled by the Hubble constant at a given redshift $H(z)$, where $H^2(z) = H_0 \left(\Omega_m (1 + z)^3 + \Omega_\Lambda \right)$ such that it is dimensionless. The shear tensor $\Sigma_{ij}$ has been defined with a minus sign so as to make its largest eigenvalue correspond to the fastest collapsing axis. The trace-free component of $\Sigma_{ij}$ causes a shear while change in volume of the deformation tensor is captured by the the trace of $\Sigma_{ij}$ which corresponds to the compression (or divergence) of the velocity field:
\begin{equation}
Tr(\Sigma_{ij}) = -\frac{\nabla \cdot \vec{v}}{H(z)}
\end{equation}
The eigenvalues and corresponding eigenvectors (\eone, \etwo~and \ethree) of the velocity shear tensor are obtained at each grid cell. The eigenvalues describe the strength of the expansion or collapse of matter along the corresponding eigenvectors. For example, three large positive eigenvalues indicate strong collapse along all three principal axes. The eigenvectors of the shear tensor, evaluated at the position of each host halo, provide an orthonormal ``eigenframe'' within which the anisotropy of mass infall onto halos can be naturally examined. The location where subhalos cross host's virial radius is computed in this eigenframe, namely the angle formed between the satellites position at the moment of accretion and each eigenvector, is computed. Given that the eigenvectors are non directional lines, the entry points to the virial sphere correspond to a single octant of the 3D cartesian coordinate system.

\subsection{Host halo binning}
\label{subsection:host_halo_binning}
In the previous section host halos were divided in mass bins according to $h^{-1}M_{\odot}$ in order to compute the $R_{\textrm{virmed}}$ in each mass bin and then to construct the smoothed shear field. Now we are interested in studying the accretion events of the small halos on their hosts. Therefore we bin the host halos with respect to $\tilde{M} = M_{\textrm{halo}}/M_\star(z)$ where $M_\star(z)$ is the mass of a 1 $\sigma$ peak in the density field at each redshift. It is defined by the overdensity variance within a sphere of radius $R(z) = (3M_\star(z) / 4 \pi \rho_{\textrm{crit}} ) ^{1/3}$ which is  equal to the square of the critical density threshold for spherical collapse $\delta_c^2$ \citep[e.g.][]{1997ApJ...490..493N}. 

\begin{figure}
    \begin{center}
        \begin{tabular}{c}
            \includegraphics[scale=0.65]{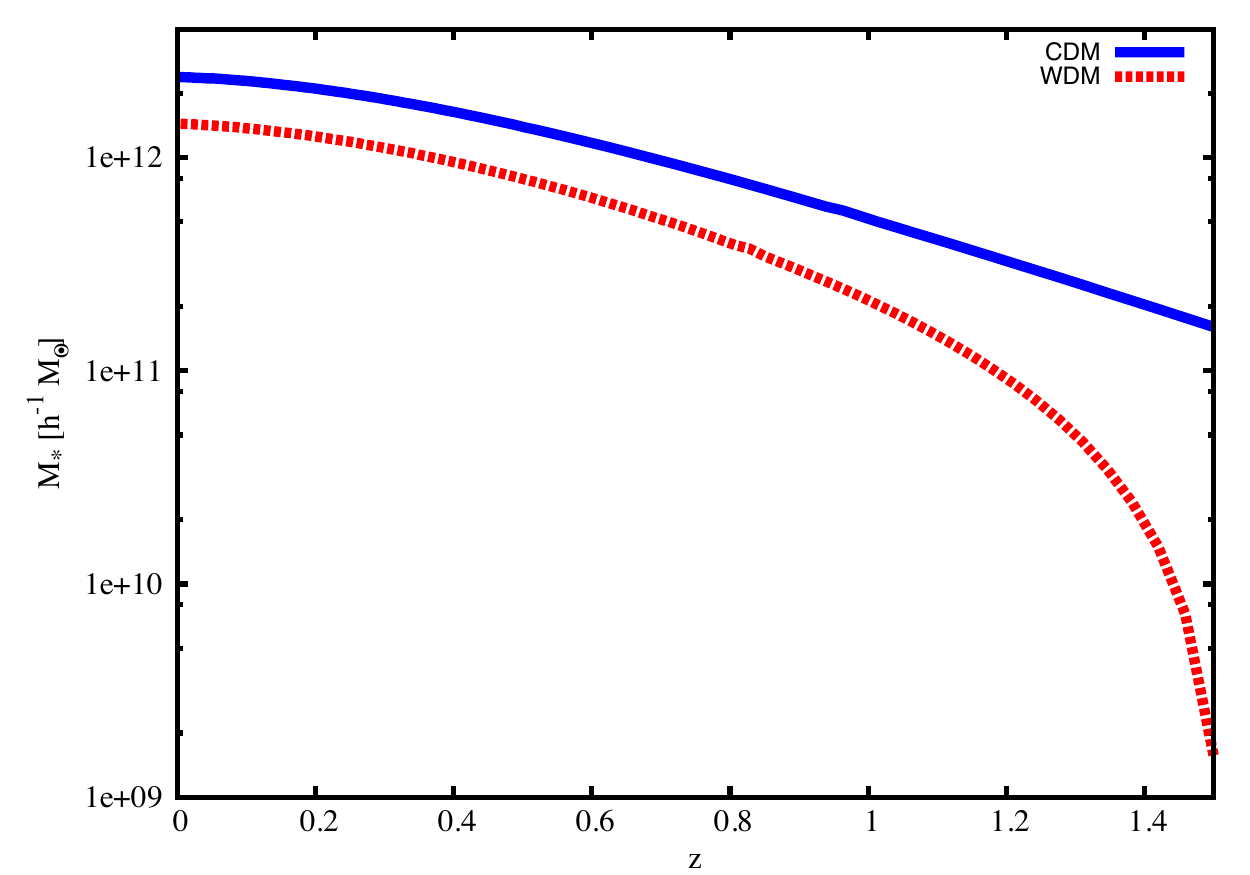}
        \end{tabular}
        \caption{A measure of the halo mass $M_\star(z)$ in units of the mass of a collapsing object at each redshift in $\Lambda$CDM (red) and $\Lambda$WDM (blue) cosmologies. $M_\star(z)$ is defined by the overdensity variance within a sphere of radius $R(z) = (3M_\star(z) / 4 \pi \rho_{\textrm{crit}} ) ^{1/3}$ which should be equal to the square of the critical density threshold for spherical collapse $\delta_c^2$. Due to the exponential drop of the power spectrum $P(k)$ in the $\Lambda$WDM cosmology the mass variance $\sigma(M)$ is below the crititical overdensity $\delta_c$ for $z > 2$ and the $M_\star(z)$ cannot be calculated using the overdensity variance. {\label{fig:Mstar}}}
    \end{center}
\end{figure}

\begin{figure}
    \begin{center}
        \begin{tabular}{c}
            \includegraphics[scale=0.46]{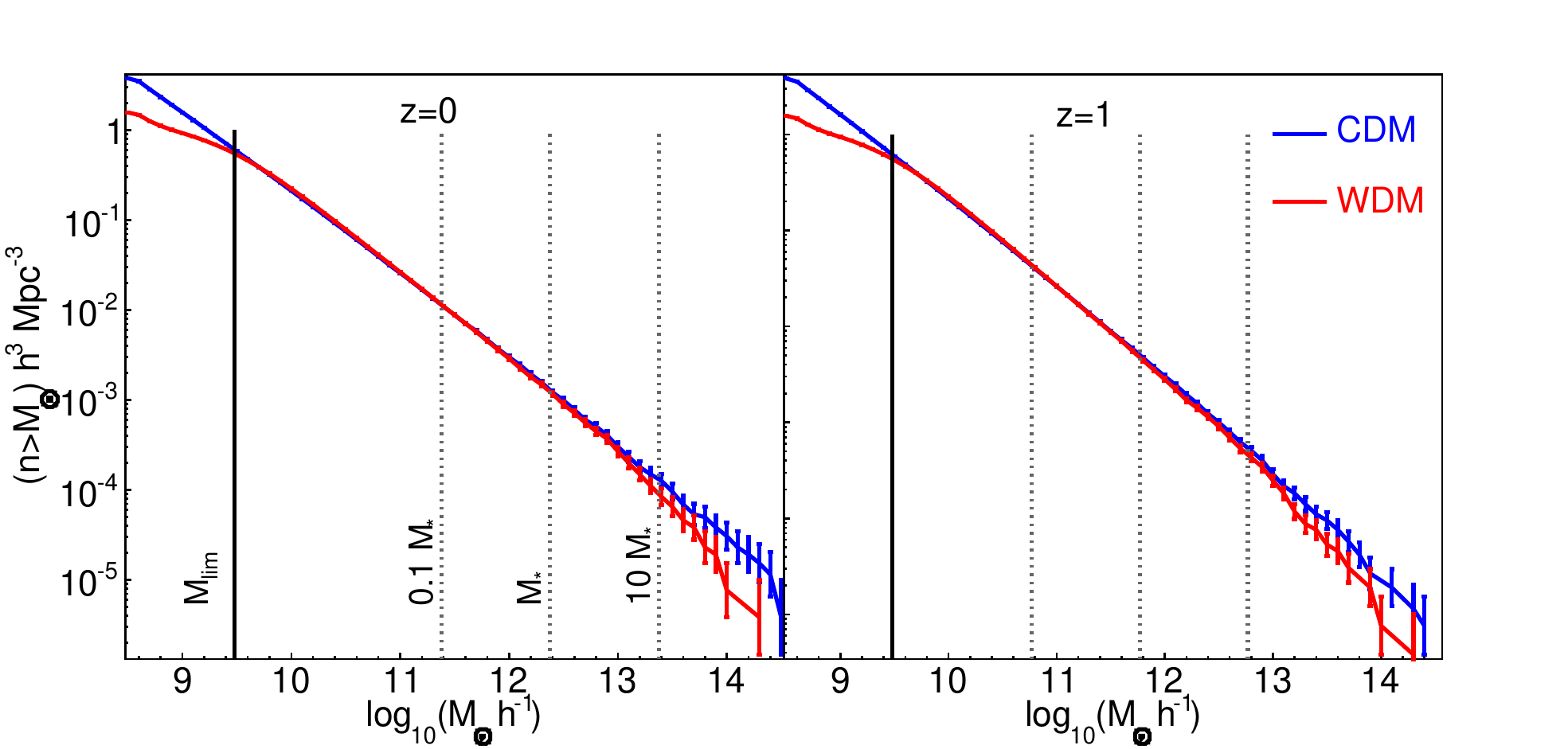}
        \end{tabular}
        \caption{Differential mass functions for the $\Lambda$CDM (blue line) and $\Lambda$WDM (red line) simulations. Left panel: at redshift $z=0$, right panel: at redshift $z=1$. The solid vertical line marks the value of the limiting mass in the $\Lambda$WDM simulation: $M_{\textrm{lim}} = 3\times10^9h^{-1}\textrm{M}_\odot$. The dashed vertical lines mark the redshift dependent delimiting mass of the host halo mass bins, from left to right $0.1\,M_\star$, $M_\star$ and $10\,M_\star$. {\label{fig:AHF_mass_functions}}}
    \end{center}
\end{figure}

The $M_\star(z)$ can be computed independently for $\Lambda$CDM and $\Lambda$WDM. The key feature of $\Lambda$WDM cosmology that distinguishes it from $\Lambda$CDM, is the lack of initial small-scale density fluctuations. This results in a different abundance of low-mass collapsed objects (DM halos) at a given redshift below the cut-off mass scale in $\Lambda$WDM and $\Lambda$CDM cosmologies. In addition, the formation time is also modified; $\Lambda$WDM halos form later relatively to the $\Lambda$CDM cosmology due to larger free-streaming length. Therefore the evolution of $M_\star(z)$ with redshift in $\Lambda$WDM cosmology differs from that observed in $\Lambda$CDM. Typically at a given redshift $M_\star(z)$ in $\Lambda$WDM will be lower than in $\Lambda$CDM as shown in Figure \ref{fig:Mstar}. Moreover, due to the exponential drop of the power spectrum $P(k)$ in the $\Lambda$WDM cosmology the mass variance $\sigma(M)$ is below the critical overdensity $\delta_c$ for $z > 2$, hence the $M_\star(z)$ becomes 
ill-defined. Therefore we decided to use the $M_\star(z)$ values as obtained for the $\Lambda$CDM as our fiducial mass unit when binning the halos into different mass regimes in both simulations. For illustrative purposes, we show in Figure \ref{fig:AHF_mass_functions} the halo mass functions in $\Lambda$CDM and $\Lambda$WDM simulations at redshifts $z=0$ and $z=1$ together with the corresponding $\{0.1,\,1,\,10\,\}\times M_\star(z)$ values. Such a binning of the host halos with respect to the CDM $M_\star(z)$ is done in both simulations at each redshift.

In sum: for each host all accretion events are identified at all redshifts. The angle formed between each accretion event and the three eigenvectors of the shear (computed on 4, 8, and 16 times the median virial radius of all halos, at that redshift, within a mass decade) is computed. The mass of the host is scaled by a characteristic mass $M_{\star}(z)$ from the CDM simulation, such that accretion onto huge halos is not compared with accretion onto small ones.

\section{Results}
\label{section:results} 

\subsection{Universal subhalo accretion pattern across different scales}
\label{subsection:results_WDM_smoothing_scales}

\begin{figure*}
    \begin{center}
        \begin{tabular}{c}
            \includegraphics[scale=0.25]{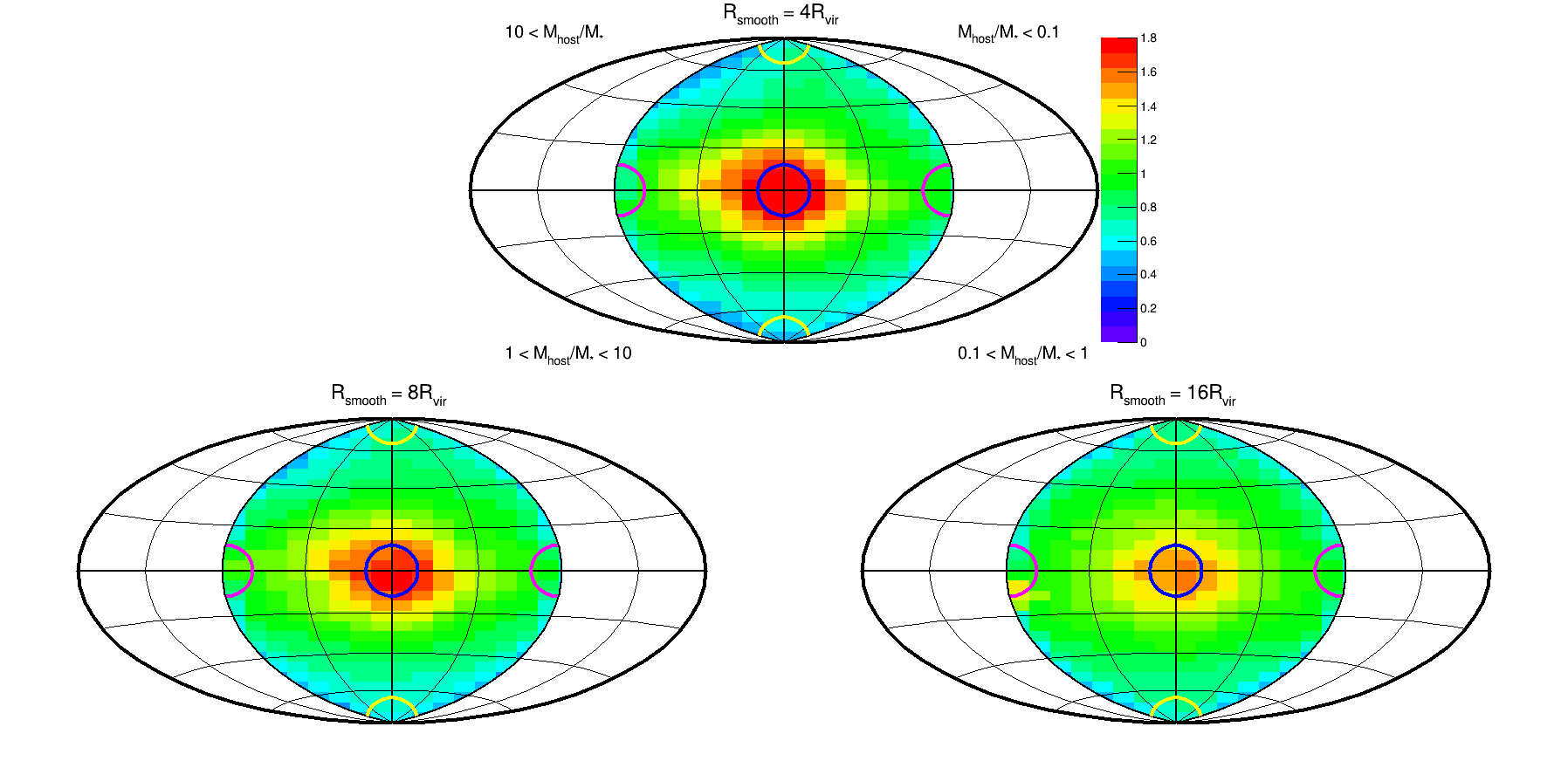}
        \end{tabular}
        \caption{The locations of subhalo entry points in $\Lambda$WDM cosmology are shown in an Aitoff projection of the virial sphere smoothed on 4 [upper], 8 [lower left] and 16 [lower right] virial radii. In each panel, starting at "noon" and going clockwise, we show these entry points for accretion events occurring onto host halos in four different mass ranges $\tilde{M}< 0.1$, $0.1 < \tilde{M} < 1$, $1 < \tilde{M} < 10$ and $10 < \tilde{M}$. $\tilde{M}$ is a measure of the halo mass in units of the mass of a collapsing object at each redshift. The density of entry points is normalized to that expected from a uniform distribution. The "north" and "south" pole correspond to \eone; the two mid points on the horizontal axis at $\pm 180^{\circ}$ to correspond to \etwo, while the midpoint corresponds to \ethree. The yellow, red and blue circles correspond to the eigenframe axes \eone,~ \etwo~and \ethree~respectively.{\label{fig:WDM_Rsmooth4_8_16}}}
    \end{center}
\end{figure*}

In Figure \ref{fig:WDM_Rsmooth4_8_16} we stack the accretion events onto all WDM host halos at all redshifts and show them in an Aitoff projection. The events correspond to the entry points to the virial sphere smoothed on 4 (top), 8 (bottom left) and 16 (bottom right) virial radii. In each panel, starting at "noon" and going clockwise, these are where $\tilde{M}< 0.1$, $0.1 < \tilde{M} < 1$, $1 < \tilde{M} < 10$ and $10 < \tilde{M}$. In essence, each quadrant of each Aitoff shows the density of accretion events as seen from the centre of the host, projected onto the virial sphere. In order to quantify the statistical significance of any anisotropy in the angular entry-point distribution, we divide the number of entry points in a given area on the virial sphere by the number of entries expected if the accretion events were uniformly distributed. The reference frame defined by the eigenvectors we plot yellow, red and blue circles around the axes defined by \eone,~ \etwo~and \ethree~respectively.

We observe a strong tendency for the accretion to occur along \ethree~regardless of the host halo mass. The effect is greatest for the most massive host halos and progressively weakens as host mass decreases. Also the beaming along \ethree~weakens with increasing smoothing, as large smoothings effectively homogenizes the LSS, randomizing the principal direction of the shear tensor.

In order to quantify the accretion anisotropy and to compare the anisotropy strength between the two considered simulations we show in Figure \ref{fig:WDM_Rsmooth4_8_16_probability_distribution} the probability distribution of the cosine of the angle formed between the subhalo entry point \racc~and the eigenvectors of the shear \eone, \etwo~and \ethree, namely $\mu_i$ = \racc$\cdot{\bf e}_{i}$, where i $\in\{1, 2, 3\}$, for the WDM and CDM simulations. Uniform distributions would be represented by a line at unity.  The strength of the alignment, $\sigma$, is characterized by the average offset between a given probability distribution and a random (uniform) one, calculated in units of the Poisson error. The average difference between the number of entry points found in a given bin and the number expected from a uniform distribution is calculated in terms of the Poisson error of a random distribution of the same size. If $\sigma$ is around unity then, on average, the measured alignment lies within the Poisson 
error of a uniform distribution while high values of $\sigma$ indicate a strong deviation from uniformity. 

The probability distributions of the three cosine angles confirm that the beaming of subhalos is strongest when the velocity shear tensor is computed on the smallest scales, and when accretion onto the largest hosts is considered in both DM scenarios. Note that at all smoothing lengths and for all the host masses the subhalos are statistically aligned with \ethree~and away from \eone~while the alignment of the infall points with \etwo~is statistically insignificant. 

In general, independently of the applied smoothing, the beaming along \ethree~and the flattening along \eone~are slightly lower in the $\Lambda$CDM than in the $\Lambda$WDM cosmology and the statistical significance of the anisotropy in $\Lambda$WDM cosmology is few (from $~\sim5$ to $\sim10\%$) percents higher than in $\Lambda$CDM. We have checked that the direct difference between the angles in $\Lambda$WDM and $\Lambda$CDM is higher than the error bars on the individual measurement in each model confirming the significance of the result.

\begin{figure*}
    \begin{center}
        \begin{tabular}{c}
            \includegraphics[scale=0.95]{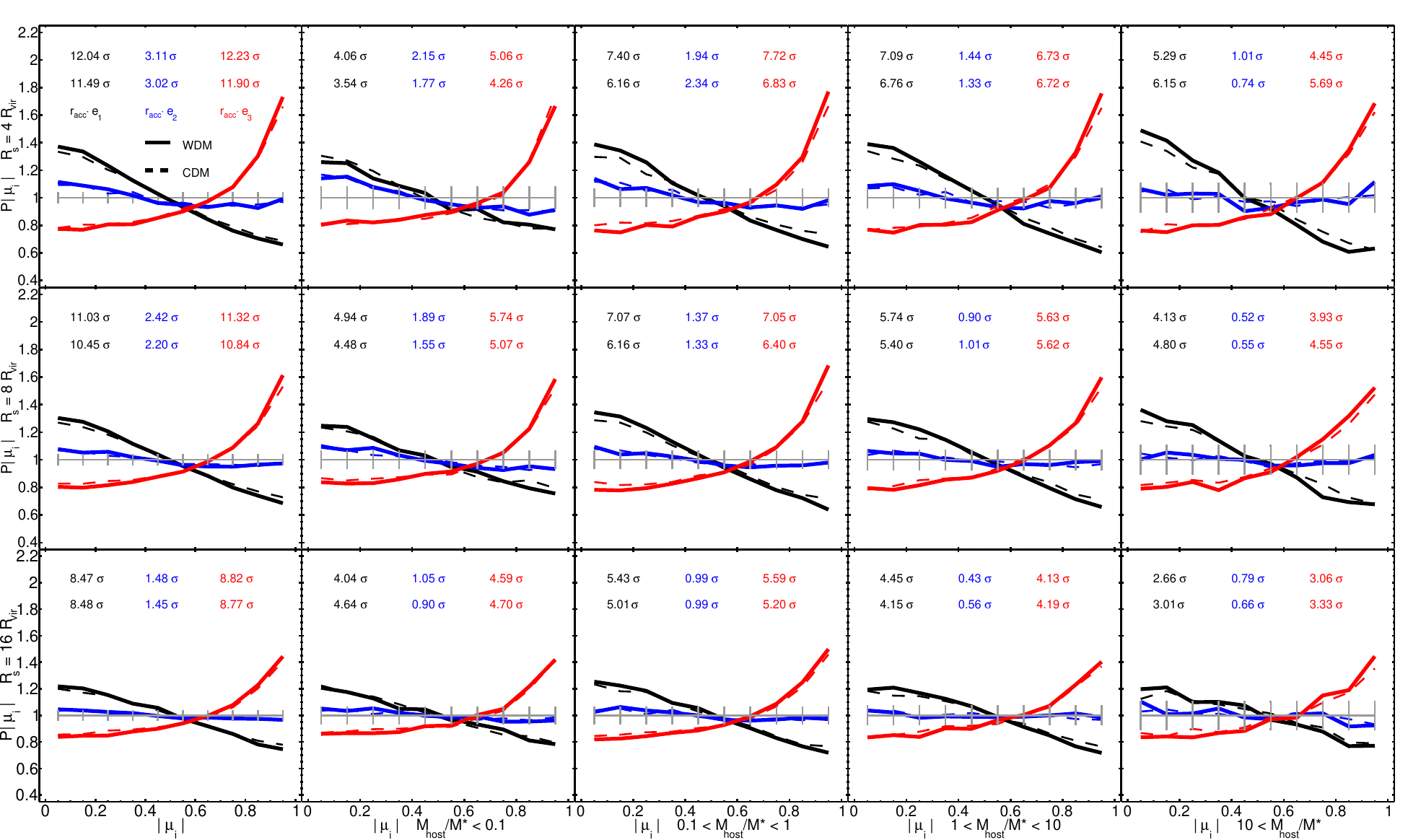}
        \end{tabular}
        \caption{The anisotropic accretion quantified by means of a probability distribution of the cosine of the angle made between a subhalo's entry point and the eigenvectors \eone~(black), \etwo~(blue) and \ethree~(red). $\Lambda$WDM cosmology is plotted as solid lines, $\Lambda$CDM - as dashed lines. The bottom, middle and top rows show the probability distribution when the shear has been smoothed on 4, 8 and 16 virial radii respectively. The probability distributions are split according to the value of the host halo mass (in the plots in the left column all the host halos are considered). The statistical significance of each probability distribution is characterized by the average offset between it and a random distribution in units of the Poisson error and is indicated by the corresponding colored number in each panel (upper line for $\Lambda$WDM, bottom line for $\Lambda$CDM cosmology). Distributions that are consistent with random have values $\sigma~\sim$ 1. {\label{fig:WDM_Rsmooth4_8_16_
probability_distribution}}}
    \end{center}
\end{figure*}

\subsection{Merger mass dependence of the anisotropic infall pattern}
\label{subsection:results_WDM_merger_mass_fcn} 

\begin{figure*}
    \begin{center}
        \begin{tabular}{c}
            \includegraphics[scale=0.25]{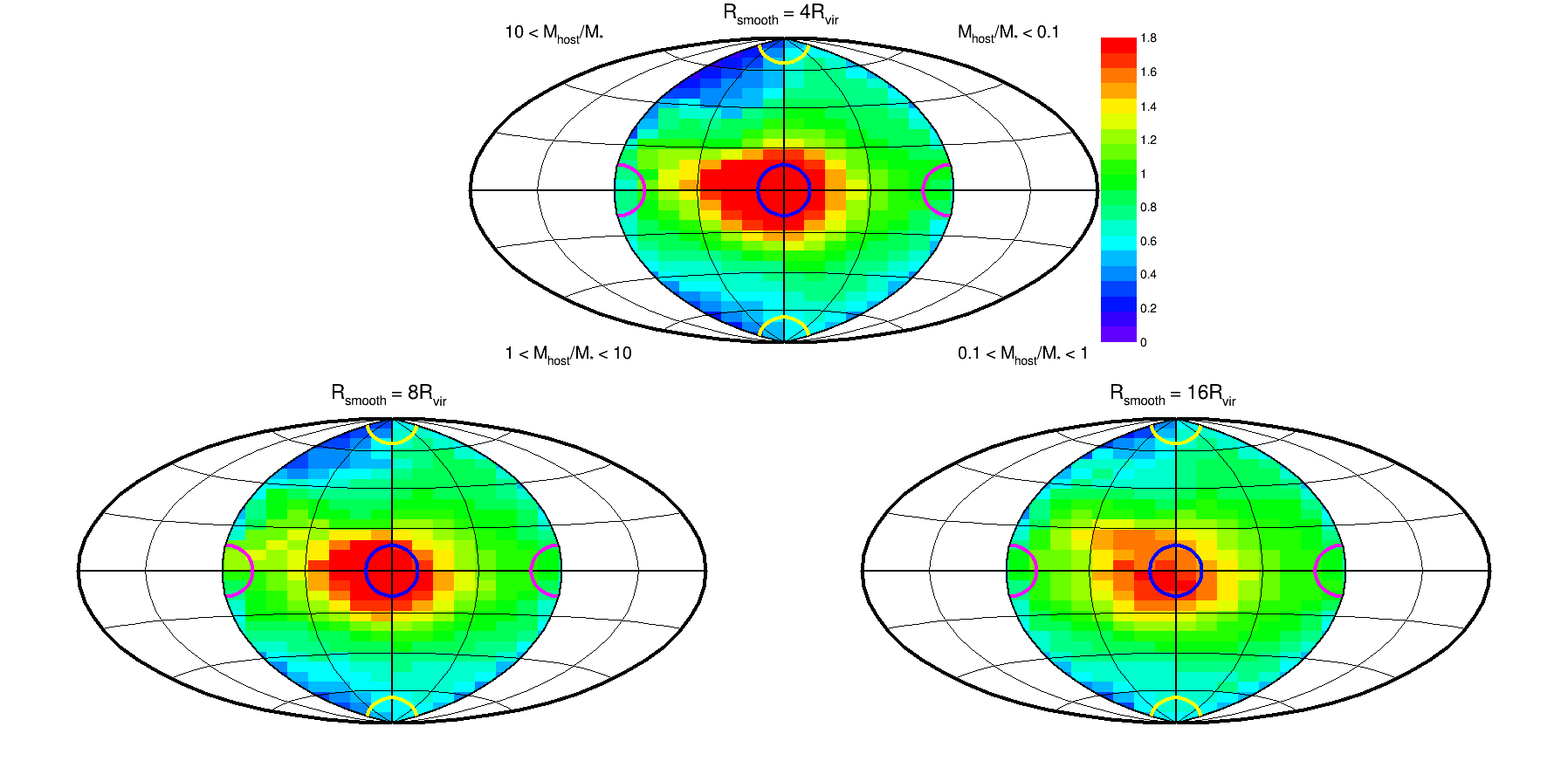}
        \end{tabular}
        \caption{The locations of subhalo entry points in $\Lambda$WDM cosmology are shown in an Aitoff projection of the virial sphere smoothed on 4 [upper], 8 [lower left] and 16 [lower right] virial radii. Only mergers whose mass exceeds 5\% of the host mass are considered. {\label{fig:WDM_Rsmooth4_8_16_Mergers02}}}
    \end{center}
\end{figure*}

Now we are interested in the infall anisotropy as function of merger mass. In Figure \ref{fig:WDM_Rsmooth4_8_16_Mergers02} we show the WDM accretion events onto host halos only for mergers whose mass exceeds 5\% of its host halo. A visual comparison with Figure \ref{fig:WDM_Rsmooth4_8_16} shows that the tendency to be accreted along \ethree~is larger for the "massive" subhalos than it was found while considering all the mergers stacked together. 

\begin{figure*}
    \begin{center}
        \begin{tabular}{c}
            \includegraphics[scale=0.6]{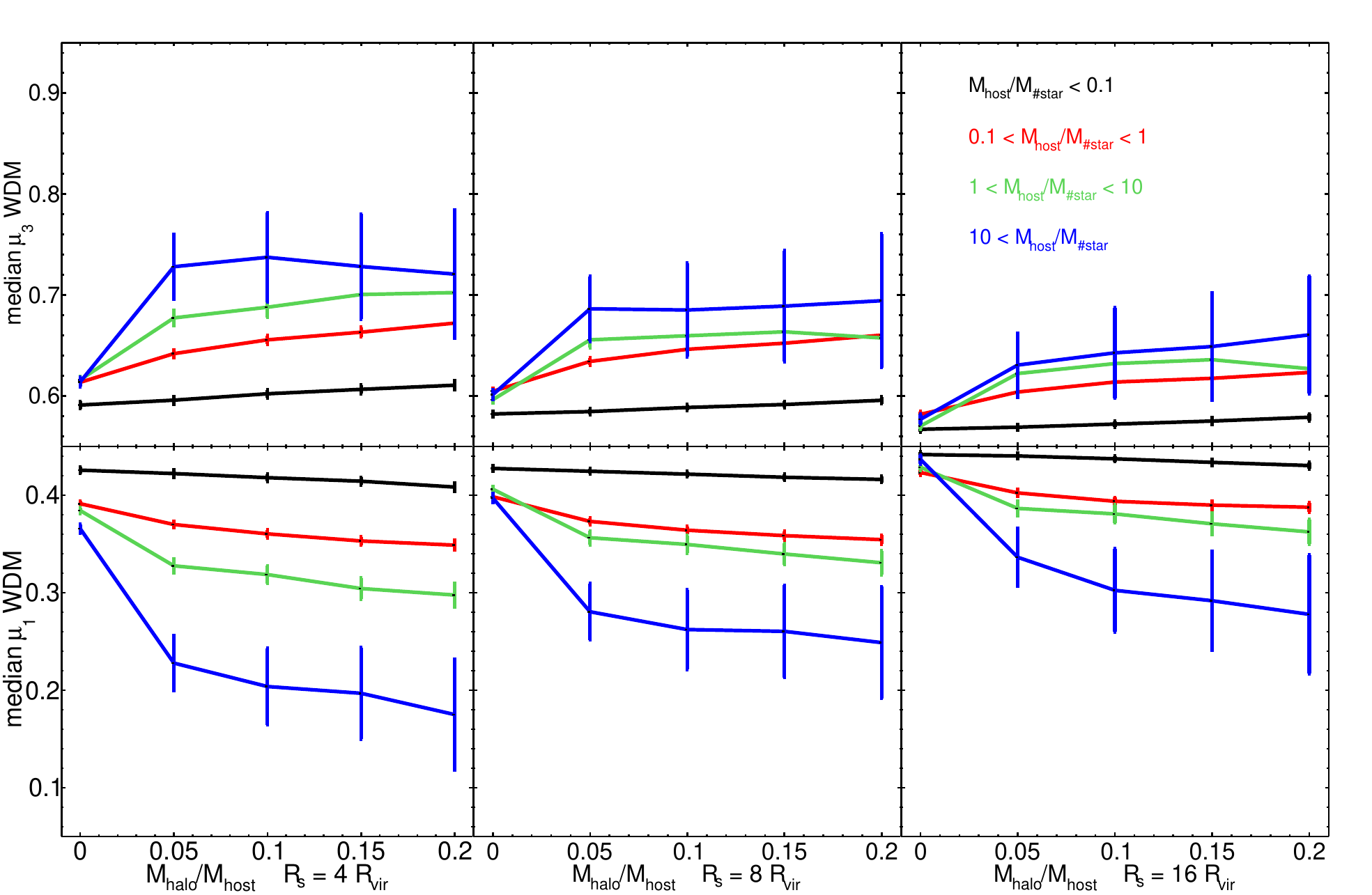}
        \end{tabular}
        \caption{Median cosine of the angle formed between the vector pointing from the center of the virial sphere to the merger accretion point and the axes defined by \eone [bottom] and \ethree [top] as the function of merger ratio $x$ defined by $M_{\rm merger} = x M_{\rm host}$. In each panel the host halos are divided into four mass bins: $\tilde{M}< 0.1$ (black), $0.1 < \tilde{M} < 1$ (red dashed), $1 < \tilde{M} < 10$ (green dotted) and $10 < \tilde{M}$ (blue dash-dotted).{\label{fig:WDM_Rsmooth4_8_16_Mergers_comparison}}}
    \end{center}
\end{figure*}

To quantify the dependence of the infall orientation on the merger mass we show in Figure \ref{fig:WDM_Rsmooth4_8_16_Mergers_comparison} the median cosines $\mu_{i}$ as function of merger ratio. At each point $x$ on the horizontal axis only mergers whose mass exceeds $x\%$ of their host halo mass are taken into account in computing the median cosines $\mu_i$.  The $3\sigma$ error bars at each point are evaluated from 10\,000 uniform distributions of the same size as the sample distribution. The median cosine is computed for each uniform distribution. These 10\,000 medians (all close to 0.5) are sorted and the spread of these values is examined by computing the 0.15 and 99.85 percentiles. This 3$\sigma$ spread is then used to get a feeling for how one would expect $\mu$ to fluctuate based solely on the sample size. Note the range of the $y$-axis 
and how far from the median value of 0.5 these curves are.

The universality of the accretion pattern is confirmed across all the mergers masses. Regardless of the merger to host mass ratio and the smoothing used, there is a statistically significant tendency for subhalos to be accreted closer to \ethree~than to any other of the eigenvectors (scale independence preserved and strengthened with increasing merger ratio). As the host halo mass and/or the merger mass increase the beaming along \ethree~gets stronger with a simultaneous flattening along \eone. This tendency is preserved independently of the smoothing scale, however with increasing smoothing the structures are less beamed and less flattened approaching to a spherically isotropic infall.

\begin{figure*}
    \begin{center}
        \begin{tabular}{c}
            \includegraphics[scale=0.6]{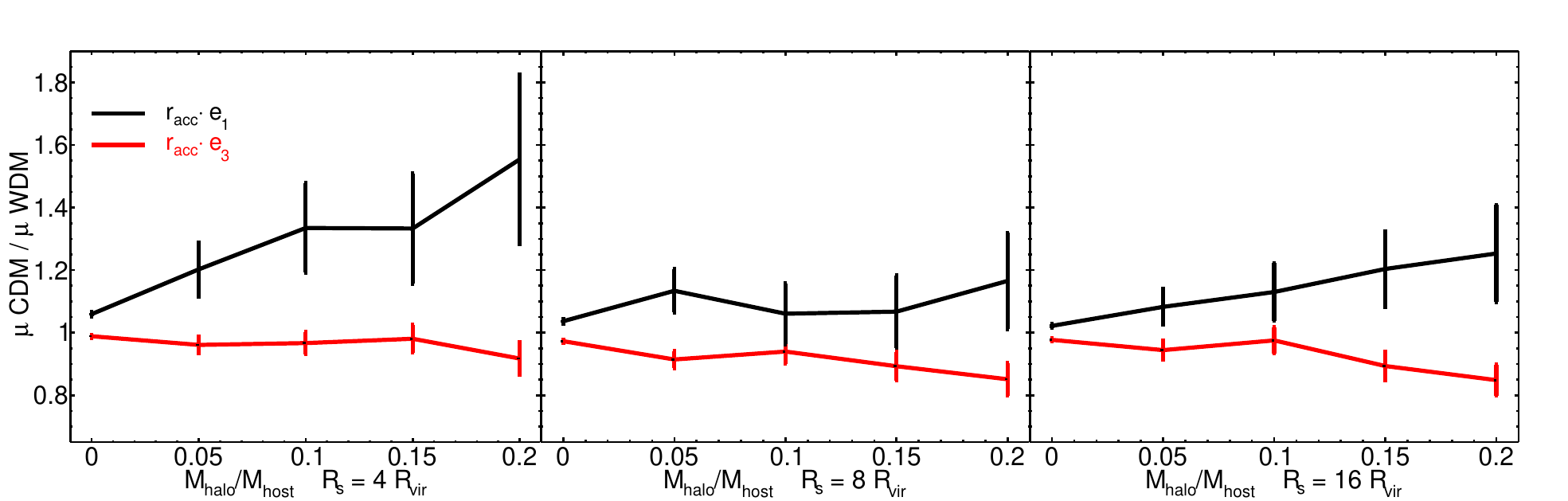}
        \end{tabular}
        \caption{Ratio of the median cosines of the accretion angles formed with \eone [black] and \ethree [red] the as the function of merger ratio $x$ defined by $M_{\rm merger} > x M_{\rm host}$. Only the accretion events onto the most massive host halos ($10 < \tilde{M}$) are shown for illustrative purposes. The enhanced beaming along \ethree~and the flattening along \eone~in $\Lambda$WDM versus $\Lambda$CDM cosmology is confirmed for all the merger mass scales and the difference $\Lambda$WDM versus $\Lambda$CDM in the flattening along \eone~is shown to be more important for heavy mergers at all smoothing scales.{\label{fig:n1_n3_in_merger_bins}}}
    \end{center}
\end{figure*}

In order to compare the merger mass dependence in $\Lambda$CDM and $\Lambda$WDM, in Figure \ref{fig:n1_n3_in_merger_bins} the $\Lambda$CDM to $\Lambda$WDM ratios of the median cosines $\mu_{1}$ and $\mu_{3}$ are plotted as function of merger ratio. Although we compared all the mass bins, for illustrative purposes only the accretion events onto the most massive host halos, namely with $\tilde{M} > 10$ are shown here. The enhanced beaming along \ethree~and the flattening along \eone~in $\Lambda$WDM as compared to $\Lambda$CDM cosmology is confirmed for all the merger mass scales. Moreover, the difference in the flattening along \eone~is more important for heavy mergers at all smoothing scales ($\mu_{1\,CDM}/\mu_{1\,WDM} \approx 1.5$ for $M_{\textrm{halo}}/M_{\textrm{host}} > 0.2$ compared to $\mu_{1\,CDM}/\mu_{1\,WDM} \approx 1.2$ for $M_{\textrm{halo}}/M_{\textrm{host}} > 0.05$ at the smoothing scale $R_{s} = 4R_{\textrm{vir}}$). The difference in the funneling along \ethree~ does not depend so strongly on 
the merger to host mass ratio but a mild increase in the $\mu_{3\,CDM}/\mu_{3\,WDM}$ ratio with increasing $M_{\textrm{halo}}/M_{\textrm{host}}$ ratio is present independently on the smoothing scale.


\subsection{Redshift dependence of the anisotropic infall pattern}
\label{subsection:results_z_dependence}

\begin{figure*}
    \begin{center}
        \begin{tabular}{c}
            \includegraphics[scale=0.75]{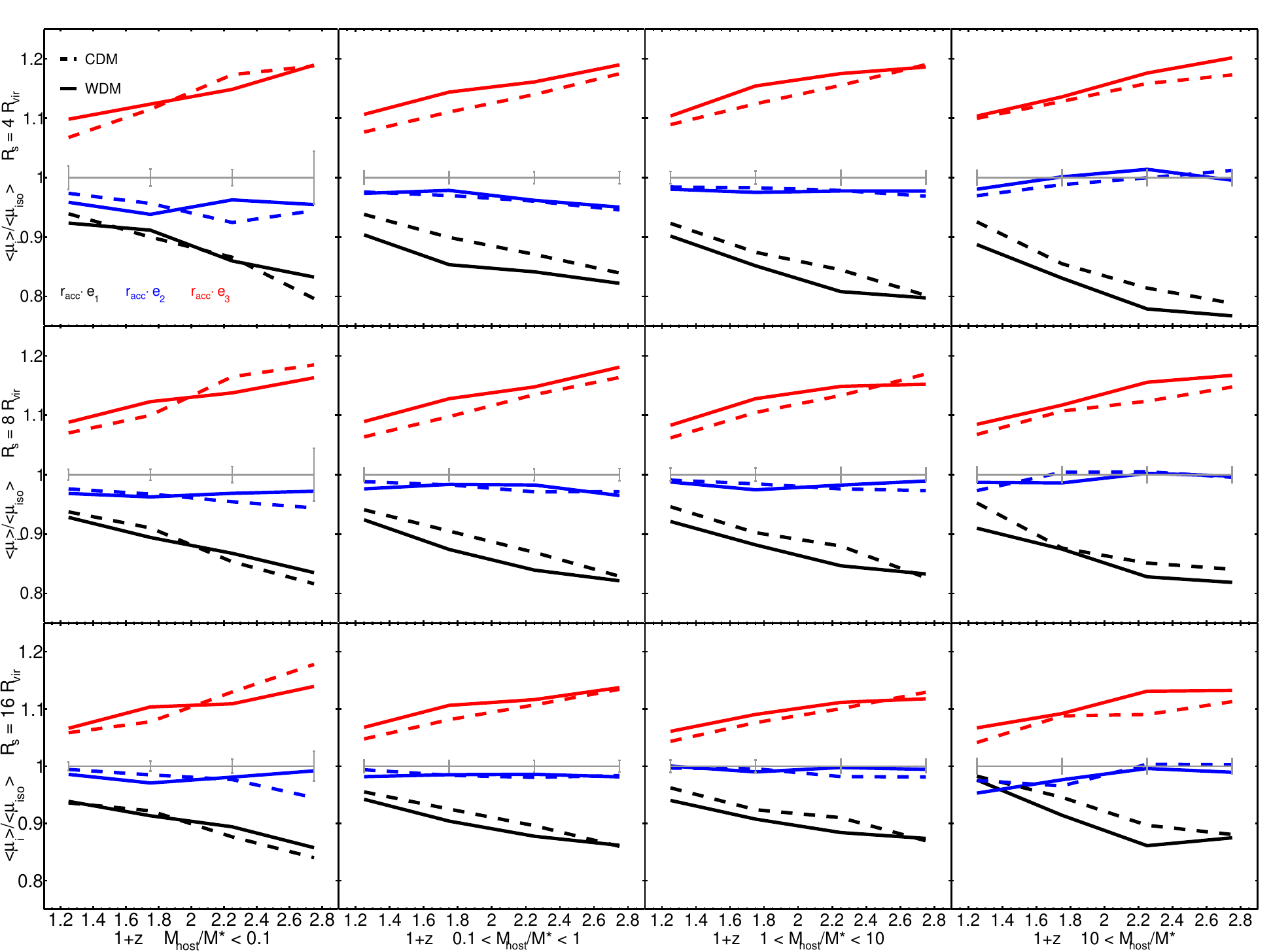}
        \end{tabular}
        \caption{The mean cosine angles, \eone (black), \etwo (blue) and \ethree (red), divided by the mean of the isotropic distribution are plotted for $\Lambda$WDM and $\Lambda$CDM cosmologies (solid and dashed lines respectively). The top, middle and bottom rows show the mean cosines ratio for the shear been smoothed on 4, 8 and 16 virial radii respectively. In each row the results are split according to the value of the host halo mass, from the left to the right: $\tilde{M}< 0.1$, $0.1 < \tilde{M} < 1$, $1 < \tilde{M} < 10$ and $10 < \tilde{M}$. {\label{fig:n1_n2_n3_vs_z}}}
    \end{center}
\end{figure*}

In previous sections we stacked together the accretion events at all redshifts and looked only at host and merging halos mass dependence. However the beaming of substructures is redshift dependent as shown in Figure \ref{fig:n1_n2_n3_vs_z} where the beaming across redshifts 0 to 2 in the $\Lambda$CDM and $\Lambda$WDM cosmologies normalized to the isotropic infall is depicted.

In both simulations the funneling along \ethree~appears across all the redshifts, however at early times it is more pronounced than at low redshifts. Consequently, the infall pattern gets more flattened along \eone~with increasing redshift while the distribution of \etwo~is consistent with the isotropic distribution across all the history of structure growth. 

We quantify the statistical probability of this infall pattern as compared to the spherically isotropic accretion by constructing 10\,000 uniform distributions at each redshift. The size of the uniform distributions is equal to the number of accretion events recorded in the $\Lambda$CDM simulation at that redshift.  The $3\sigma$ error at each redshift is constructed in the same way as in section \ref{subsection:results_WDM_merger_mass_fcn}, namely by examining the spread in 10\,000 medians of each random trial. The error bars increase in size at higher redshifts since the number of accretion events decreases due to the simulation's resolution. 

As already mentioned in section \ref{subsection:results_WDM_merger_mass_fcn}, and confirmed here at each redshift independently, the small systematic difference in the strength of the beaming along \ethree~and flattening along \eone~axes between the $\Lambda$WDM and $\Lambda$CDM cosmologies is present. We do not see any redshift dependence in the difference of the beaming between $\Lambda$WDM and $\Lambda$CDM cosmologies. The offset $\mu_{1\,CDM} - \mu_{1\,WDM}$ and $\mu_{3\,CDM} - \mu_{3\,WDM}$ is consistent with a constant given the statistical error.

We note that the differences between $\Lambda$CDM and $\Lambda$WDM are systematic but not terribly significant. Although we see trends wherein subhalos in $\Lambda$WDM are more strongly beamed towards the most massive hosts along \ethree~at all redshifts, we readily concede that such trends are within the expected poissonian error. Statistically speaking the geometry of accretion events is the same in $\Lambda$WDM or (above the limiting mass $M_{\rm lim}$) in $\Lambda$CDM.

\subsection{Delayed evolution of the anisotropic accretion pattern in $\Lambda$WDM}
\label{subsection:results_merger_mass_evolution_with_redshift} 

The typical mass of a halo, both the host halo and the infalling one, at a given redshift is higher in $\Lambda$CDM than in $\Lambda$WDM scenario due to the lack of the initial small-scale density fluctuations and later halo formation in $\Lambda$WDM. On the other hand, as stated in section \ref{subsection:results_z_dependence}, the beaming along \ethree~and flattening along \eone~axes is stronger in $\Lambda$WDM than in $\Lambda$CDM cosmology at each redshift in the range $0<z<2$ and the anisotropy strength increases with redshift in both scenarios.

Thus, we expect that, in the same way as the earlier CDM halo formation, the strength of the beaming in $\Lambda$CDM, although weaker than the beaming in $\Lambda$WDM at the same redshift $z_1$, was of the same strength at some earlier redshift $z_2 > z_1$ when the average mass of the infalling objects was the same as in $\Lambda$WDM at $z_1$. 

\begin{figure*}
    \begin{center}
        \begin{tabular}{c}
            \includegraphics[scale=0.55]{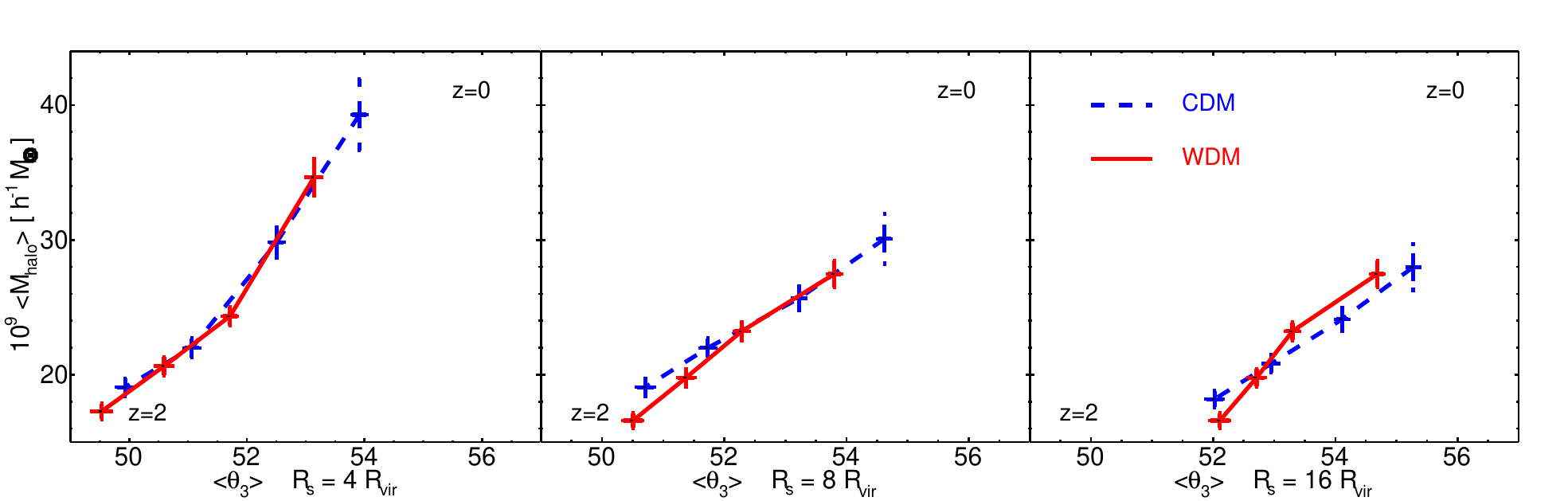}
        \end{tabular}
        \caption{Average accretion angles $\theta_3$ and average sub-halo masses in $\Lambda$CDM and $\Lambda$WDM cosmologies (blue dashed and solid red lines respectively) at four equidistant redshifts between $z=2$ and $z=0$ (bottom left corner and top right corner of each panel).{\label{fig:average_mass_average_n3_with_redshift}}}
    \end{center}
\end{figure*}

This is exactly what we show in Figure \ref{fig:average_mass_average_n3_with_redshift} where each point $(\langle \theta_3 \rangle, \langle M_{\textrm{halo}} \rangle)$ is computed at a redshift between $z=0$ and $z=2$ so that at each $z$ we have a characteristic infall angle $\langle \theta_3 \rangle$ and a characteristic infalling sub-halo mass $\langle M_{\textrm{halo}} \rangle)$ in both cosmologies. There is a clear correlation between the average infalling angle and the average sub-halo mass. The average infall angles $\langle \theta_3 \rangle_{\textrm{WDM}}(z_1)$ and $\langle \theta_3 \rangle_{\textrm{CDM}}(z_2)$ computed at redshifts $z_2$ and $z_1$ when the average sub-halo masses were equal $\langle M_{\textrm{halo}} \rangle_{\textrm{WDM}}(z_1) = \langle M_{\textrm{halo}} \rangle_{\textrm{CDM}}(z_2)$, are also equal. The analogous correlation is present between the average angle $\langle \theta_1 \rangle$ and $\langle M_{\textrm{halo}} \rangle$.

Therefore, we can state that the accretion anisotropy is fully universal in the sense that the accretion pattern is independent on the shear smoothing scale, redshift, host and sub-halo masses together with the considered cosmology ($\Lambda$CDM and $\Lambda$WDM). The beaming strength, however, is modulated with the shear smoothing scale, host and sub-halo masses and redshift. The difference in the strength of the accretion beaming along \ethree~and flattening along \eone~in $\Lambda$WDM with respect to $\Lambda$CDM, although not very significant, follows the same delayed evolution as the typical mass of the infalling objects confirming once more its universality.

\section{Summary and conclusion}
\label{section:summary}
In this paper we have focused on the merging of small halos with massive ones in the warm and cold dark matter cosmologies. We have characterized the subhalos anisotropic infall pattern with respect to the scale and redshift independent eigenframe of each host halo defined by the eigenvectors of the ambient velocity field. 

We have shown WDM subhalos (and confirmed that CDM subhalos) tend to be accreted on their hosts from a specific direction with respect to the reference frame defined by the velocity shear eigenvectors on each host halo. As halo is formed by accreting ambient (i.e. non-collapsed) material from its surrounding, the accretion proceeds first along \eone, where the collapse is fastest, and then along \etwo. At a given time after the halo virialization most of the formed substructures lie in the plane defined by \etwo~and \ethree. This is why the funneling angle with respect to \eone~is large and most of the material that crosses the virial sphere is being funneled along \ethree~- axis of slowest collapse. This is exactly what we find in this work across a wide range of redshifts and for the mergers and hosts from the highest masses formed in our simulations $\sim10^{14}h^{-1}\textrm{M}_\odot$ down the the lowest reliable mass $M_{\textrm{lim}} = 3\times10^9h^{-1}\textrm{M}_{\odot}$.

This generic infall pattern is modulated in strength depending on the chosen smoothing scale: smaller smoothing scale corresponds to stronger beaming. Also, the most massive hosts/mergers tend to accrete/be accreted more anisotropically than their lighter partners.

Moreover, the beaming of subhalos in a given host halo mass bin narrows with increasing redshift.  This may be because at high redshifts the accretion is dominated by higher mass ratio mergers. As shown, the alignment with \ethree~is most pronounced for the most massive hosts and their most massive satellites. Therefore the narrowing of accretion at high redshift may be partly due to the lack of more isotropic accretion from smaller halos contaminating and diluting the signal. This is either because they are not resolved in our simulation or because they have not collapsed yet. As suggested by \cite{1970A&A.....5...84Z}, the first structures to form via gravitational instability are anisotropic ``pancakes'', defined by their principle axis of collapse \eone. Pancakes lead to filaments as material collapses next along \etwo. This process is more extreme in the early universe. In broad terms, by the time our matter had collapsed into virialized halos, the only axis left  along which they can travel is \ethree. 

We have compared our results from a warm dark matter universe to the dynamics characteristic in the cold dark matter scenario. The anisotropic infall pattern, inferred from simulations, is found to be independent on whether the dark matter particle is cold or warm.  While the main findings are the same in both cases, we find that in the $\Lambda$WDM universe the accretion on the most massive host halos is slightly more anisotropic than the $\Lambda$CDM: the $\Lambda$WDM subhalos infall is more beamed along \ethree~and more squashed along \eone~across all the redshifts considered in this paper. However these are not strong conclusions and for the most part the beaming is the same in the two cosmologies. 

The main difference between the two models is the abundance of small halos and sub-halos and the modified halo formation time. We have shown that the alignment of accretion of the small sub-halos, and the accretion onto small halos, tends to dilute the alignment with \ethree. Hence, fewer number of small halos and sub-halos in the $\Lambda$WDM cosmology, compared to $\Lambda$CDM at a given redshift, implies stronger alignment with \ethree~in the former case. We have shown that the strength of the alignments of the infall directions is correlated with the mass of the infalling objects and thus follows the history of halo formation. The later halo formation in $\Lambda$WDM cosmology is reflected in later "homogenization" of the accretion beaming with respect to $\Lambda$CDM.

It is known that halos shape and spin are correlated with the velocity shear eigenvectors and the strength of the alignment depends on the halo mass (\citealt{2013MNRAS.428.2489L}). Thus it would be interesting to compare the degree of correlation and the strength of these alignments between the $\Lambda$CDM and $\Lambda$WDM cosmologies.

On the other hand, a natural following step would be to include gas physics to the simulations. Gas filaments are known to be narrower than the DM filaments, thus we can expect to see a higher level of anisotropy in the distribution of accreted gas by the halos. Furthermore, the transmission of the angular momentum from one parcel of gas to another may be highly effective and would lead to higher homogeneity of the properties of the accreted direction, enhancing the effect of spin and shape alignment. 

Observationally, the anisotropic infall could be compared to satellite distributions relying on galactic surveys such as SDSS, JWST, Euclid or LSST. Given that these (future) surveys will provide more detailed photometry of galaxies and more precise distance measurements for thousands of objects over the whole sky, combined with the increasing power and resolution of cosmological simulations, we will be able to infer more information regarding the dynamic properties of the dark matter and galactic halos. This will undoubtedly give an insight into the nature of the dark matter particle.

\section*{Acknowledgments}

BK and HC research was conducted within the framework of the Lyon Institute of Origins under grant ANR-10-LABX-66. 

AK is supported by the {\it Ministerio de Econom\'ia y Competitividad} and the {\it Fondo Europeo de Desarrollo Regional} (MINECO/FEDER, UE) in Spain through grant AYA2015-63810-P. He also acknowledges support from the {\it Australian Research Council} (ARC) grant DP140100198. He further thanks Jacqueline Taieb for le coeur au bout des doigts.

SG and YH have been partially supported by the Deutsche Forschungsgemeinschaft under the grant GO563/21-1. 

GY acknowledges financial support from MINECO under research grants  AYA2012-31101 and AYA2015-63810.

The authors wish to thank the Red Espa\~nola de Supercomputaci\'on for granting us access to the Marenostrum supercomputer at the Barcelona Supercomputer Center, where the simulations presented in this paper have been performed.

\bibliographystyle{mnras}
\bibliography{Allrefs}

\end{document}